\documentclass[submission,copyright,creativecommons]{eptcs}

\makeatletter
\let\O@argtabularcr\@argtabularcr
\def\O@xtabularcr{\@ifnextchar[\O@argtabularcr{\ifnum 0=`{\fi}\cr}}
\let\O@tabacol\@tabacol
\let\O@tabclassiv\@tabclassiv
\let\O@tabclassz\@tabclassz
\let\O@tabarray\@tabarray
\def\author@tabular{\authorsize\def\@halignto{}\@authortable}
\let\endauthor@tabular=\endtabular
\def\author@tabcrone{{\ifnum0=`}\fi\O@xtabularcr\affilsize\itshape
 \let\\=\author@tabcrtwo\ignorespaces}
\def\author@tabcrtwo{{\ifnum0=`}\fi\O@xtabularcr[-3\p@]\affilsize\itshape
 \let\\=\author@tabcrtwo\ignorespaces}
\def\@authortable{\leavevmode \hbox \bgroup $\let\@acol\O@tabacol
 \let\@classz\O@tabclassz \let\@classiv\O@tabclassiv
 \let\\=\author@tabcrone \ignorespaces \O@tabarray}
\makeatother

\def\authorrunning{Hall, Varanasi, et al.}
\def\titlerunning{Knowledge-Assisted Reasoning with Goal-directed ASP and Event Calculus} 

\usepackage[usenames,dvipsnames]{xcolor}
\usepackage[fleqn]{amsmath}
\usepackage{algorithm}
\usepackage{algpseudocode}
\usepackage[inline]{enumitem}
\usepackage{graphicx}
\usepackage{tikz-cd}
\usepackage{multirow} 
\usepackage{cprotect}
\usepackage{comment}
\usepackage[T1]{fontenc}
\usepackage{caption}
\usepackage{diagbox}
\usepackage{listings}
\usepackage{wrapfig}
\usepackage[compact]{titlesec}
\usepackage[font=small,skip=0pt]{caption}
\definecolor{darkbrown}{rgb}{0.4, 0.26, 0.13}
\definecolor{burgundy}{rgb}{0.5, 0.0, 0.13}
\usepackage[frozencache,cachedir=.]{minted}
\usepackage{multicol}
\newcommand{\mycomment}[1]{}

 \titlespacing{\section}{2.75pt}{2.75pt}{2.75pt}
 \titlespacing{\subsection}{2pt}{2pt}{2pt}

\def\BibTeX{{\rm B\kern-.05em{\sc i\kern-.025em b}\kern-.08em
    T\kern-.1667em\lower.7ex\hbox{E}\kern-.125emX}}

\begin{document}

    \title{Knowledge-Assisted Reasoning \\of Model-Augmented System Requirements
    with Event Calculus and Goal-Directed Answer Set Programming\footnote{UT Dallas authors are supported by NSF (grants IIS 1718945, IIS 1910131, IIP 1916206), Amazon and DoD. Support for this project also comes from the FIT BUT internal project FIT-S-20-6427 and by the H2020 ECSEL project Arrowhead Tools. We would also like to dedicate our work to the memory of second author's Father.
}}


\author{
    Brendan Hall$^1$ \and
    Sarat Chandra Varanasi$^2$ \and
    Jan Fiedor$^3$ \and
    Joaqu{\'i}n Arias$^4$ \and
    Kinjal Basu$^2$ \and
    Fang Li$^2$ \and
    Devesh Bhatt$^1$ \and
    Kevin Driscoll$^1$ \and
    Elmer Salazar$^2$ \and
    Gopal Gupta$^2$
    \institute{$^1$Honeywell Advanced Technology, Plymouth, USA}
    \institute{$^2$The University of Texas at Dallas, Richardson, USA}
    \institute{$^3$Honeywell International s.r.o \& Brno Univ. of Technology, Brno, Czech Republic}
    \institute{$^4$Universidad Rey Juan Carlos, Madrid, Spain} 
    \email{$^{1,3}$\{brendan.hall, jan.fiedor, devesh.bhatt, kevin.driscoll\}@honeywell.com}
    \email{$^2$\{sarat-chandra.varanasi, kinjal.basu, fang.li, ees101020, gupta\}@utdallas.edu}
    \email{$^4$joaquin.arias@urjc.es}
}

\maketitle

\begin{abstract}
We consider requirements for cyber-physical systems represented in constrained natural language. We present novel automated techniques for aiding in the development of these requirements so that they are consistent and can withstand perceived failures.  We show how cyber-physical systems' requirements can be modeled using the \textit{event calculus} (EC), a formalism used in AI for representing actions and change. We also show how answer set programming (ASP) and its query-driven implementation s(CASP) can be used to directly realize the event calculus model of the requirements. This event calculus model can be used to automatically validate the requirements. Since ASP is an expressive knowledge representation language, it can also be used to represent contextual knowledge about cyber-physical systems, which, in turn, can be used to find gaps in their requirements specifications. We illustrate our approach through an altitude alerting system from the avionics domain.
\end{abstract}


   \lstset{%
       escapeinside={(*}{*)},%
      }
 



\section{Introduction}

Developing effective requirements is crucial for success in building a system. 
The earlier the requirements are validated, the fewer problems system developers will encounter later. Current automation of requirements engineering tasks attempt to ensure their consistency and adequacy, namely, that they can withstand perceived failures. However, such automated support remains limited. In this work, we present novel automated techniques for aiding the development of model-augmented requirements that are adequate---to the extent that adequacy can be established---and consistent. Thus, we can have more confidence in the requirements. We limit ourselves to requirements for cyber-physical systems, particularly those in avionics. We assume that requirements are generated within the MIDAS (Model-Assisted Decomposition and Specification) \cite{hall2020model} environment, and are expressed in CLEAR (Constrained Language Enhanced Approach to Requirements)~\cite{hall2018clear}, a constraint natural requirement language based on EARS \cite{mavin2009easy}.

Our main contribution in this work is to show how the \textit{Event Calculus} (EC)~\cite{sergot1986logic,shanahan1999event} and  \textit{Answer Set Programming} (ASP) \cite{gelfond2014knowledge} can be used to formalize constrained natural language requirements for cyber-physical systems and perform knowledge-assisted reasoning over them.  ASP is a logic-based knowledge representation language that has been prominently used in AI. Our work builds upon recent advances made within the s(CASP) system~\cite{arias2018constraint}, a query-driven (or goal-directed) implementation of predicate ASP that supports constraint solving over reals, permitting the faithful representation of time as a continuous quantity. The s(CASP) system permits the modeling of event calculus directly~\cite{arias2019modeling-short}.  
An advantage of using the event calculus---in contrast to automata and Kripke structure-based approaches---is that it can directly model cyber-physical systems, thereby avoiding ``pollution" due to (often premature) design decisions that must be made otherwise.
The event calculus is a formalism---a set of axioms---for modeling dynamic systems and was proposed by artificial intelligence researchers to solve the \textit{frame problem}~\cite{sergot1986logic, shanahan1999event}.
Our methods have been developed within the MIDAS framework of Honeywell \cite{hall2020model}. 
The MIDAS framework embodies the essence of the \textit{Property Management Model} \cite{micouin2014model} within a model-augmented requirement ecosystem, which integrates constrained natural language requirements within a layered information model expressed in Object Process Model (OPM)~\cite{dori2011object}. 
The details of MIDAS, PMM and OPM are not important here, instead, the primary goal of this work is to explore how constrained natural language requirements,  specified within MIDAS,  using the CLEAR notation, can be automatically reasoned about and analyzed using the event calculus and query-driven answer set programming. Specifically, we explore:


\begin{enumerate}
  \item{How ASP-based model checking (over dense time) can validate specified system behaviors wrt system properties.}
  \item{How application of \textit{abductive reasoning} can extend ASP-based model checking to incorporate domain knowledge and real-world/environmental assumptions/concerns.}
  \item{How knowledge-driven analysis can identify typical requirement specification errors, and/or requirement constructs which exhibit areas of potential/probable risk.}
 \end{enumerate}


The rest of the paper is organized as follows. 
Section \ref{enabling} introduces the enabling technologies of answer set programming (ASP), a goal-directed implementation of ASP called s(CASP), and the event calculus. Section \ref{clear} discusses the CLEAR notation for specifying requirements, and illustrates it with an example. Section \ref{demo-case-study} presents an application developed within Honeywell Corporation using our approach. 
We also illustrate requirement defect discovery using s(CASP) for property-based model-checking as well as discuss how more general knowledge of potential requirements defects may detect defects that traditional techniques may not be able to find. 
Section \ref{concl} presents our conclusions and future work.

The main contribution of our research is to show how the event calculus and goal-directed answer set programming together serve as a promising framework for modeling and reasoning over cyber-physical and avionics systems' requirements. We also demonstrate the utility of abductive reasoning to find the assumptions under which a property holds or does not hold (along with automatically providing justification in both cases) for requirements specification. Given that ASP is a knowledge representation language, we also illustrate how knowledge about system context (e.g., processing platform  may malfunction causing them to be reset) can be utilized to automatically find gaps in requirements as well.

\section{Enabling Technologies}
\label{enabling}

\subsection{Answer Set Programming}

Answer Set Programming (ASP) is a declarative paradigm that provides the stable model semantics for logic programs (with negation-as-failure). ASP is a highly expressive paradigm that can elegantly express complex reasoning methods, including those used by humans, such as default reasoning, deductive and abductive reasoning, counterfactual reasoning, constraint satisfaction, etc. ~\cite{baral,gelfond2014knowledge}.

\mycomment{ASP supports better semantics for negation ({\it negation as failure}) than does standard logic programming and Prolog. An ASP program consists of rules that look like Prolog rules. The semantics of an ASP program {$\Pi$} is given in terms of the answer sets of the program \texttt{ground($\Pi$)}, where \texttt{ground($\Pi$)} is the program obtained from the substitution of elements of the \textit{Herbrand universe} for variables in $\Pi$~\cite{baral}.}
The rules in an ASP program are of the form:
  \texttt{p :- q$_1$, ..., q$_m$, not r$_1$, ..., not r$_n$.}
where $m \geq 0$ and $n \geq 0$. Each of \texttt{p} and \texttt{q$_i$} ($\forall i \leq m$) is a literal, and each \texttt{not r$_j$} ($\forall j \leq n$) is a \textit{naf-literal} (\texttt{not} is a logical connective called \textit{negation-as-failure} or \textit{default negation}). The literal \texttt{not r$_j$} is true if proof of {\tt r$_j$} \textit{fails}. Negation as failure allows us to take actions that are predicated on failure of a proof. Thus, the rule {\tt r :- not s.} states that {\tt r} can be inferred if we fail to prove {\tt s}. Note that in the rule above, {\tt p} is optional. Such a headless rule is called a constraint, which states that conjunction of {\tt q$_i$}'s and \texttt{not r$_j$}'s should yield \textit{false}. Thus, the constraint {\tt :- u, v.} states that {\tt u} and {\tt v} cannot be both true simultaneously in any model of the program (a model is called an answer set).

\mycomment{The declarative semantics of an Answer Set Program $\Pi$  is given via the Gelfond-Lifschitz transform~\cite{baral,gelfond2014knowledge} in terms of the answer sets of the program \texttt{ground($\Pi$)}. ASP also supports classical negation. A classically negated predicate (denoted {\tt -p}) means that {\tt p} is definitely false. Its definition is no different from a positive predicate, in that explicit rules have to be given to establish {\tt -p}. More details on ASP can be found elsewhere~\cite{baral,gelfond2014knowledge}. 

The goal in ASP is to compute an {\it answer set} given an answer set program, i.e., compute the set that contains all propositions that if set to true will serve as a model of the program (those propositions that are not in the set are assumed to be false).  Intuitively, the rule above says that {\tt p} is in the answer set if {\tt q$_1$, ..., q$_m$} are in the answer set and {\tt r$_1$, ..., r$_n$} are not in the answer set. }
ASP also allows for assumption-based reasoning via \textit{abduction} \cite{harman65}. Abduction is a form of reasoning concerned with the generation and evaluation of explanatory hypotheses. In abductive reasoning, given the premise $P \Rightarrow Q$, and the observation $Q$, one surmises or \textit{assumes} (\textit{abduces}) that $P$ holds. More generally, given a theory $T$, an observation $O$, and a set of abducibles $A$, then $E$ is an explanation of $O$ (where $E \subset A$) if:

~~~~1. $T \cup E \models O$

~~~~2. $T \cup E$ is consistent 

\noindent We can think of abducibles $A$ as a set of assumptions. Abduction allows us to find ``gaps" in our knowledge that must entail a property in order for that property to hold. 

\subsection{The s(CASP) System}


Considerable research has been done on answer set programming since the inception of the stable model semantics that underlies it~\cite{gelfond2014knowledge}. A major problem with ASP implementations is that programs have to be grounded and SAT-solver-based implementations such as CLASP~\cite{clingo} used to execute the propositionalized program to find the answer sets. There are multiple problems with this SAT-based implementation approach, which include exponential blowup in program size, having to compute the entire model, and not being able to  produce a justification for a conclusion. Goal-directed implementations of ASP~\cite{arias2018constraint}, called s(ASP) and s(CASP), work directly on predicate ASP programs (i.e., no grounding is needed) and are query-driven (similar to Prolog). The s(ASP) and s(CASP) systems only explore the parts of the knowledge-base that are needed to answer the query, and they provide a proof tree that serves as justification for the query. The s(ASP) and s(CASP) systems support predicates with arbitrary terms as arguments as well as constructive negation \cite{arias2018constraint}.
 
In the work reported here, we will mainly use the s(CASP) system that additionally supports constraint solving over reals, which is important for reasoning faithfully about (continuous) time. The s(CASP) system is the key technology for representing and analyzing CLEAR requirements modeled with the event calculus. The s(CASP) system also directly supports abductive reasoning and can provide justification for a query.

\subsection{Event Calculus}
 
The Event Calculus~\cite{sergot1986logic,shanahan1999event} is a formalism for modeling dynamic systems which originated from the field of AI where it was devised to solve the \textit{frame problem}. 
The event calculus is organized around events and fluents. 
We use the event calculus formalism to model an evolving world in a cyber-physical system. The EC models the world that is changing due to \textit{events} that are happening that, in turn, influence \textit{fluents}, where a fluent is a variable that changes with time such as location of an object or temperature of the boiler. EC provides axioms for stating the conditions under which events \textit{happen} or conditions under which events \textit{initiate}, \textit{terminate}, or \textit{release} a fluent. EC also provides axioms for when a fluent \textit{holds}. EC models the laws of inertia that hold in the real world. For instance, the location (a fluent) of an object does not change unless the event ``move'' happens. Event calculus uses reasoning over time, as events are occurrences in time.  The Event Calculus is elegantly modeled using ASP, in particular, with the query-driven s(CASP) system due to its support for real-time constraints.

The basic event calculus axioms are shown in Fig.~\ref{fig:BEC-axioms}. The \texttt {happens/2} predicate models the occurrence of an event $e$ at a particular time instance $t_i$, while \texttt{holdsAt/2} specifies the condition under which a fluent $f$ holds at time $t_j$. The axioms are easy to follow.\mycomment{, however, we give a brief explanation below.} More details can be found elsewhere~\cite{mueller_book,arias2019modeling-short}.
\mycomment{
\begin{description}
\item[Axiom BEC1.] A fluent $f$ is stopped between time points $t_1$ and
$t_2$ iff it is terminated or released by some event $e$ that occurs after
$t_1$ and before $t_2$.
\item[Axiom BEC2.] A fluent $f$ is started between time points $t_1$ and
$t_2$ iff it is initiated or released by some event $e$ that occurs after
$t_1$ and before $t_2$.
\item[Axiom BEC3] A fluent $f_2$ is true at time  $t_2$ if
$Trajectory(f_1,t_1,f_2,t_2)$ is true, fluent $f_1$ is initiated by an event $e$
 at time  $t_1$, and $f_1$ is not stopped between $t_1$ and
$t_2$.
\item[Axiom BEC4.] A fluent $f$ is true at time $t$ if it is true at
time 0 and is not stopped between time 0 and $t$.
\item[Axiom BEC5.] A fluent $f$ is false at time $t$ if it is false at
time 0 and it is not started between time 0 and $t$.
\item[Axiom BEC6.] A fluent $f$ is true at time $t_2$ if it is initiated
by some event $e$ occurring at some earlier time $t_1$ and it is not
stopped between $t_1$ and $t_2$.
\item[Axiom BEC7.] A fluent $f$ is false at time $t_2$ if it is
terminated by some event $e$ occurring at some earlier time $t_1$ and it is
not started between $t_1$ and $t_2$.
\end{description}
}

In a cyber-physical system, events correspond to actuator actions and fluents correspond to effects of actuators. Fluents can also describe the conditions under which events can occur.  A fluent is an entity whose value may change over time and, thus, in case of a CPS, they reflect changes in sensor readings as well as internal component(s) state. Fluents may be impacted by events, e.g, furnace shutoff (event) may lead to temperature (fluent) to drop. \textit{The state of a system is represented by the values of all its fluents.} The event calculus simply models how events lead to changes in fluents. Fluents, in turn, may trigger events, e.g., temperature reaching 80 degrees may cause furnace shutoff. Thus, \textit{one can model a cyber physical system directly in the event calculus without having to make any representational assumptions}. This is in contrast to other formalisms such as automata-based or Kripke structure-based approaches where one has to first discern the states of the system followed by the transitions between these states and then model it. The discerning of states and transitions for these methods is done too early in the design process leading to (possibly erroneous) assumptions being incorporated into the design. \textit{In the event calculus, states are implicit, as a state is defined as consisting of values of all fluents at any given moment}. \textit{Because of the implicit nature of system states, a state needs to be constructed only at the time we wish to verify something, e.g., a property.}

\begin{figure*}
\newcommand{\Axiom}[3]{\textbf{#1} & #2 &\\[0.5em]
  \multicolumn{3}{r}{#3} \\[0.75em]}
\newcommand{\AxiomB}[3]{\textbf{#1}  #2  #3\\[0.5em]}
\centering
{
  \small
  \begin{tabular}{p{16cm}}
    \AxiomB{BEC1.}%
    {$StoppedIn(t_1,f,t_2)  \equiv $}%
    {$\exists\ e,t\   (Happens(e,t) \land\ t_1<t<t_2 \land (Terminates(e,f,t) \lor Releases(e,f,t)))$} 
    \AxiomB{BEC2.}%
    {$StartedIn(t_1, f, t_2)  \equiv $}%
    {$\exists\ e, t  (Happens(e,t) \land\ t_1 < t < t_2 \land (Initiates(e,f, t) \lor Releases(e, f, t)))$}
    \AxiomB{BEC3.}%
    {$HoldsAt(f_2, t_2) \leftarrow$}%
    {$Happens(e, t_1) \land Initiates(e, f_1, t_1) \land Trajectory(f_1, t_1, f_2, t_2) \land \lnot StoppedIn(t_1, f_1, t_2)$} 
    \AxiomB{BEC4.}%
    {$HoldsAt(f, t)  \leftarrow$}%
    {$ InitiallyP(f ) \land \lnot StoppedIn(0, f, t)$}
    \AxiomB{BEC5.}%
    {$\lnot HoldsAt(f, t) \leftarrow$}%
    {$InitiallyN(f ) \land \lnot StartedIn(0, f, t)$}
    \AxiomB{BEC6.}%
    {$HoldsAt(f, t_2) \leftarrow$}%
    {$ Happens(e, t_1) \land Initiates(e, f, t_1) \land\ t_1 < t_2 \land \lnot StoppedIn(t_1, f, t_2)$}
    \AxiomB{BEC7.}%
    {$\lnot HoldsAt(f, t_2)   \leftarrow  $}%
  {$ Happens(e, t_1) \land Terminates(e, f, t_1) \land\ t_1 < t_2 \land \lnot StartedIn(t_1, f, t_2)$}
  \end{tabular}
}
  \caption{Formalization of BEC (Basic Event Calculus) axioms~\protect\cite{mueller_book}.}
\label{fig:BEC-axioms}
\end{figure*}

\section{CLEAR Notation for Requirements}
\label{clear}

Constrained Language Enhanced Approach to Requirements (CLEAR) extends the Easy Approach to Requirements Syntax (EARS) developed initially by Alistair Mavin~\cite{mavin2009easy}. EARS is an industrially pragmatic approach based on using five structured templates and keywords. Studies \cite{mavin2016listens,mavin2010big} have shown that use of EARS reduces requirements errors while improving requirement quality and readability. The EARS keywords and templates are illustrated in Figure~\ref{fig:ears}.

Although the EARS approach is often presented as a syntactic guide to writing requirements, we have found the power of EARS to be more related to the \textit{mindset} than  \textit{syntax}. The different perspectives that are explored using EARS-based keywords stimulate a lot of discussion relating to the core intent of the requirement under consideration. The separation and clarification of event and state semantics, using the \texttt{WHILE} and \texttt{WHEN} constructs are useful to constrain and define the behavioral details of cyber-physical systems at different levels of abstraction.

The EARS approach also calls for the separation of the nominal and off-nominal specification and the \texttt{IF} keyword is reserved for defining the required response to an unwanted or exceptional condition.   In practice we have found that it is often the insufficient understanding and naive assumptions regarding system failure scenarios that lead to omissions of critical requirements. 

EARS designates optional and configurable behaviors/attributes with the \texttt{WHERE} keyword. Once again, this is also an area of potential vulnerability regarding requirement completeness.
Significant hazards can emerge from misconfiguration, as illustrated by the A400M incident in Seville~\cite{a400mseville}. Hence, separating concerns helps address such issues in a more systematic fashion.

CLEAR embodies the full spirit of EARS  and refines the terms used within EARS constructs to improve specificity. CLEAR also introduces additional constructs such as  \textit{Upon Initialization} to force a systematic examination of known problem areas, such as system initialization.
CLEAR is used today to support automated analyses (e.g. requirement consistency checking) and software level test generation~\cite{bhatt2018clearfide}.   Through MIDAS, it is intended to move and automate such analysis to the system level.

\subsection{Mapping Requirements to the Event Calculus}

\begin{figure}[ht]
\begin{center}
  \includegraphics[width=0.9\textwidth]{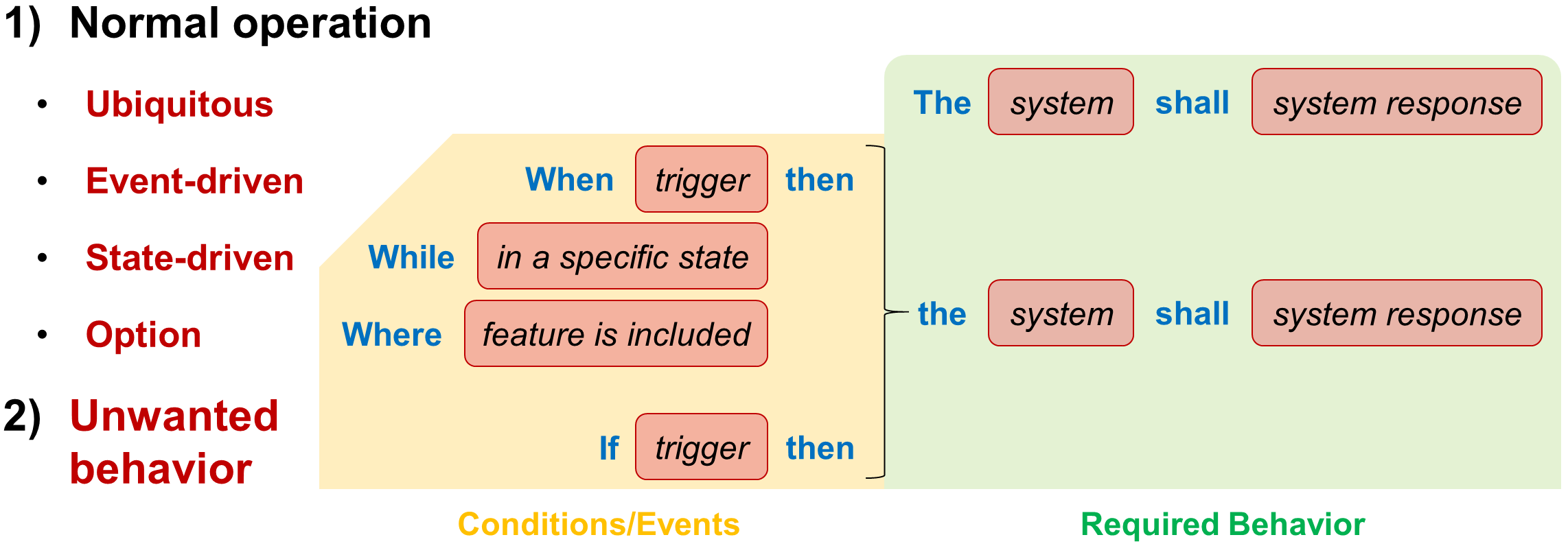}
  \vspace{1em}
  \caption{EARS Templates}
  \label{fig:ears}
\end{center}
\end{figure}

We discuss the mapping of CLEAR-based requirements \cite{hall2018clear} to the Extended Event Calculus formalization of Shanahan\cite{shanahan1999event}. CLEAR is a notation for expressing requirements for cyber physical systems based on EARS (Easy Approach to Requirement Syntax) widely adopted in the avionics industry. For simplicity, we limit ourselves to discussing only EARS. EARS provides natural language sentence templates for an engineer specifying a system to follow for writing requirements. Keywords `WHEN', `WHERE' and `IF'-`THEN' are used in these templates and play a major role. 
We make use of the extended event calculus formalism because it enables events to have duration. For cyber-physical real-time systems, response times are important,  hence formally budgeting the allocation of time throughout the levels of function \& temporal decomposition are primary concerns. Figure \ref{fig:ears} summarizes the EARS notation. Details about EARS can be found elsewhere \cite{mavin2009easy,mavin2010big}.

\vspace{-0.4cm}

Conceptually, the mapping of the extended event calculus to CLEAR is relatively straightforward, given that EARS already informally  separates event and state semantics.  To formalize the mapping, we only need to formally bound the level of temporal abstraction over which the CLEAR requirements are defined.  The EARS keyword \texttt{WHEN} then needs to define the requirements in relation to the \textit{upward} or \textit{downward} perspective, with the downward facing view having the form of state invariant constraints, and the upward facing view declaring actionable behaviours. 
For actionable behaviours,
the level of temporal abstraction also guides which of the EARS keywords to apply to the specification as outlined below:
\begin{itemize}
    \item \textbf{WHEN}: discrete event-driven specification is used when a single observation at the level of temporal abstraction is sufficient to verify the required behavior. That is, \texttt{WHEN} establishes causal relationship.
    \item \textbf{WHILE}: continuous state-driven specification is used when multiple observations at the level of temporal abstraction are needed to  verify the required behavior. {\tt WHILE} may also be used as a state qualifier for event-driven specifications, in cases where there are stateful conditions guarding the event.  
\end{itemize}

\noindent Within MIDAS, functional influence manifests as the stateful changes to objects that surround the functional intent boundary.  This maps very cleanly to the extended event calculus, once the state of the said objects at the function boundary are encoded as \textit{fluents} within the event calculus formalism.  Hence, to map the MIDAS CLEAR actionable requirements to the extended event calculus, we simply perform the following (details of MIDAS are not important here):
\begin{enumerate}
    \item Declare each MIDAS object state as a fluent.
    \item Declare {\tt happens} predicate that characterize the discrete changes in state that triggers the {\tt WHEN} condition. 
    \item Declare \texttt{initiates} and \texttt{terminates} predicates to bind the happens predicates to the influenced external state changes, with \texttt{holdsAt} state qualifiers for all {\tt WHILE} preconditions .
    \item Declare \texttt{trajectory} predicates to characterize continuous changes that evolve over multiple intervals of the temporal precision.
\end{enumerate}

\noindent We illustrate the mapping of an EARS requirement related to aircraft design:
\begin{quote}
  \textbf{While} the \textcolor{red}{aircraft is on-ground}, \textbf{when} the \textcolor{blue}{requested door position becomes open},  the door control system \textbf{shall} \textcolor{OliveGreen}{change the state of the cargo door from closed to open} \textbf{within} \textcolor{violet}{10 secs}.
\end{quote}

\noindent Mapping this to the EARS templates (Fig. \ref{fig:ears}) we see that the aircraft is on-ground is a \textcolor{red}{precondition}, the requesting of the cargo door to open is the \textcolor{blue}{trigger}, and the opening of the cargo door within 10 seconds is the \textbf{system response}, requiring the \textcolor{OliveGreen}{change of environmental state}, with an assumed level of \textcolor{violet}{temporal precision}.
In this requirement the functional intent manifests as a change in the physical state of the cargo door, in  response to a change in state that reflects the pilot requesting the door to open.
Hence, the event calculus formalism of this requirement is as follows:

\vspace{-0.3cm}
\small
\medskip\noindent
\begin{minted}[linenos,xleftmargin=.5cm]{scasp.py:SCASPLexer -x }
fluent(door_requested_position(RP)). % Map object states to fluents
fluent(door_state(DS)).
fluent(aircraft_state(AS)).
happens(pilot_requests_door_to_open, T) :- % Map change of state trigger to happens
    -holdsAt(door_requested_position(open), T1), 
    holdsAt(door_requested_position(open), T2), T2 #=< T1+10, T1 #< T, T #=< T2.
initiates(pilot_requests_door_to_open, door_state(open), T) :-
    holdsAt(aircraft_state(is_on_ground), T). % Map changes influenced by event
terminates(pilot_requests_door_to_open, door_state(closed),T) :- 
    holdsAt(aircraft_state(on_ground), T). % Map changes influenced by event
door_response_is_correct_condition :- % A typical query verifying door behaves correctly
    -holdsAt(door_requested_position(open), TB), -holdsAt(door_state(open), TB),
    holdsAt(aircraft_state(on_ground), TH), holdsAt(door_state(open), TE), 
    TB #=< TH, TB #< TE, TE #=< TB+10.
\end{minted}
\normalsize

\noindent  According  to the semantics of the  Extended Event Calculus \cite{shanahan1999event}, a fluent is considered to be initiated at any time point during the event duration. This is semantically consistent with the additional \textit{within} constraint that the CLEAR requirement notation \cite{hall2018clear} provides. Note also that the extended event calculus formalism does not distinguish  between the sensing of the environmental change and the resulting time of establishing environmental influence.  This is also consistent with the level of functional and temporal decomposition at the current level of abstraction.  Hence, when validating and discharging properties and constraints above this  level of specification, 
the event calculus model will explore the impact of the function influence starting at any point within the event duration, which in our example is 10 seconds.

Note also that the initiated action, i.e., the changing of the door state, is guarded by the aircraft state of `on ground'. Should this hold when the respective triggering event happens, the initiates predicate will fail, and therefore there will be no external functional influence.
The mapping of IF is the same as the {\tt WHEN}, given that they are both conditioned on events. The primary difference is that IF characterizes triggering off-nominal state transitions such as component failure scenarios.

As noted above actionable {\tt WHILE} requirements are used to characterize continuous behaviours and/or constraints that need to be expressed over multiple intervals at the current level of temporal abstraction. They map to trajectories within the extended event calculus formalism. We show an example and its mapping to EC below:-

\begin{quote}
\textbf{While} the \textcolor{red}{door is opening}, the \textcolor{OliveGreen}{rate of change in door position} shall \textcolor{blue}{increase positively} with a  maximum rate of \textcolor{violet}{5 degrees per second}. 
\end{quote}


\small
\begin{minted}[linenos,xleftmargin=.5cm]{scasp.py:SCASPLexer -x}
trajectory(door_opening,T1,door_position(B),T2) :- 
        holdsAt(door_position(A),T1), 
        holdsAt(door_position(B),T2), A #< B, B #=< A+5, T2 #= T1+1.
door_rate_ok :-  % Constraint on rate at which door opens is true throughout
        trajectory(door_opening,T1,door_position(P),T2), 
        holdsAt(door_opening,TH), T1 #=< TH, TH #=< T2. 
\end{minted}
\normalsize

\vspace{-0.2cm}
ASP with EC can also be used to test the adequacy of the requirements to the extent possible. In the case of avionics, the knowledge is simple and relates to Single Event Upsets (or SEUs)~\cite{normand1996single} which can induce random resets of avionic software systems. We can model the destructive power of the SEU using a \textit{reset} event. The \textit{reset} event overrides the constructed internal state of the system, forcing the initialization state to be re-established.
Our goal in introducing this reset is to move the designer to consider how robust the system initialization logic is to such transient resets.
In ASP, an extraneous event (e.g., reset) is represented as an \textit{abducible} (a reset may or may not happen). We would want to know if our system will still behave correctly in presence of this extraneous event. The knowledge assumes that the reset event can only override the cyber system (software state) and does not affect the continuous state of the physical world. Whether our system will behave correctly or not can be established with the abductive reasoning supported by s(CASP). This form of abductive reasoning has significant potential in helping establish the completeness of  requirements for cyber-physical systems.



\section{Demonstration Case Study}
\label{demo-case-study}

To explore the feasibility of our approach, we applied it to several systems under development at Honeywell. The primary goal of these case-studies is to explore how specialist domain knowledge can be incorporated into the requirement review process to help assess requirement sufficiency. Rather than encode hard-coded knowledge of specific cases, we are striving to illustrate the application of more general reusable knowledge. Our findings have been encouraging, and we have been able to supply useful feedback to live engineering teams. We illustrate our ideas with one of the systems developed.

\subsection{Altitude Alerting System}
\begin{figure*}[h]
\begin{center}
  \includegraphics[width=0.85\textwidth]{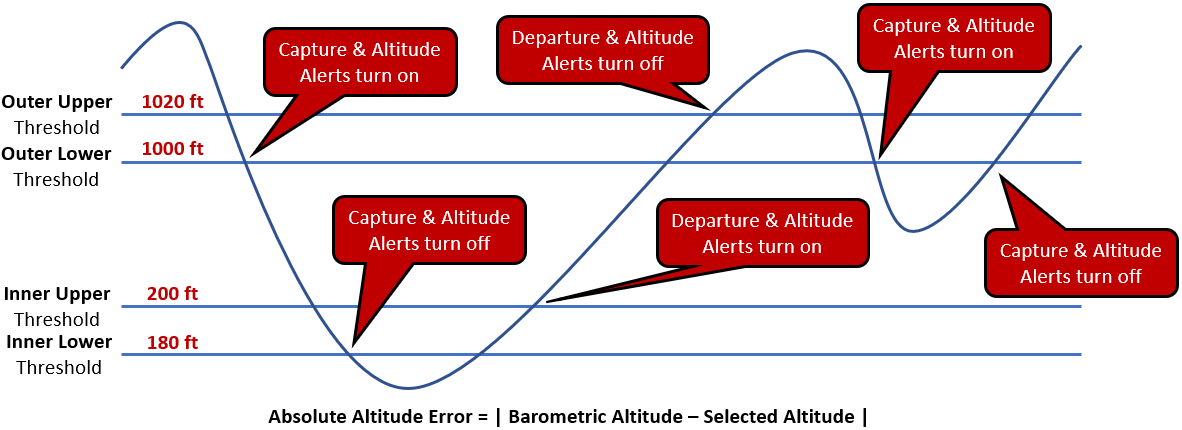}
  \caption{Altitude Alerts w.r.t Altitude Error Thresholds}
  \label{fig:altitude}
\end{center}
\end{figure*}

The altitude alerting system should issue alerts about aircraft altitude error thresholds in a timely fashion. Figure \ref{fig:altitude} shows the alerts that should be turned on/off within one second as the absolute altitude crosses alert thresholds. We first present the various notions associated with the altitude alert system's requirements, and show their mapping from CLEAR to Event Calculus and ASP

\noindent During the operation of the aircraft, the pilot  selects a target altitude. The alert system monitors the aircraft's barometric altitude trajectory and issues alerts when this barometric altitude becomes potentially hazardous in relation to the selected altitude. If the error crosses certain threshold values, then the system should promptly warn the pilot so that she can take appropriate action. Similarly, the alerting system can also withdraw the alerts if other threshold values are met. The two primary alerts are \textit{capture alert} and \textit{departure alert} explained in Figure \ref{fig:altitude}. In the following, {\tt lt} and {\tt gt} are abbreviated for `less than' and `greater than', respectively. The predicate {\tt error\_becomes\_lt(1000,T)} asserts if the altitude error becomes \textit{less than} 1000 ft. at time {\tt T}.
\newcommand{\clearreq}[1]{\textit{#1}}
\newcommand{\clearvar}[1]{\textbf{#1}}
\newcommand{\clearstate}[1]{'\textbf{#1}'}

%


\smallskip

\noindent\textbf{REQ1:}\footnote{It should be noted that these requirements are significantly simpler than the original internal baseline which consisted of a state-based specification. This simplification was enabled by MIDAS' conceptual modeling approach, which forces the specification to be expressed using conceptual abstractions, over object states that surround the boundary of intent.}
 \textit{When the \clearvar{absolute\_altitude\_error}\footnote{absolute\_altitude\_error is a derived fluent} becomes less than 1000 ft, the Altitude Alerting System shall initiate the \clearvar{capture\_alert} and \clearvar{altitude\_alert} within 1 second.}

\noindent The set of constraints, in s(CASP), in the body of \mintinline{scasp.py:SCASPLexer -x}{happens(init_cap_alert)} capture the notion that the event has to occur within one second of the error dropping below 1000 ft. This shows the ease of expressing continuous real-valued constraints in s(CASP).


\vspace{-0.15cm}
\small
\begin{multicols}{2}
\begin{minted}[linenos,xleftmargin=.5cm,autogobble]{scasp.py:SCASPLexer -x}
happens(init_cap_alert,T) :- 
error_becomes_lt(1000,T1),T1#<T,T#<T1+1.
happens(init_alt_alert,T):- 
error_becomes_lt(1000,T1),T1#<T,T#<T1+1.
initiates(init_cap_alert,cap_alert_on,T). 
terminates(init_cap_alert,cap_alert_off,T). 
initiates(init_alt_alert,alt_alert_on,T). 
terminates(init_alt_alert,alt_alert_off,T). 
\end{minted}
\end{multicols}
\vspace{-0.15cm}
\normalsize

\smallskip
\noindent\textbf{REQ2:}  \clearreq{When the \clearvar{absolute\_altitude\_error}  becomes greater than 200 ft, the Altitude Alerting System shall initiate the \clearvar{departure\_alert} and \clearvar{altitude\_alert} within 1 second.}

\vspace{-0.15cm}
\small
\begin{multicols}{2}
\begin{minted}[linenos,xleftmargin=.5cm]{scasp.py:SCASPLexer -x}
happens(init_dep_alert,T) :- 
error_becomes_gt(200,T1),T1#<T,T#<T1+1.
initiates(init_dep_alert,dep_alert_on,T). 
terminates(init_dep_alert,dep_alert_off,T). 
happens(init_alt_alert,T) :-
error_becomes_gt(200,T1),T1#<T,T#<T1+1.
\end{minted}
\end{multicols}
\vspace{-0.15cm}
\normalsize

\noindent \textbf {REQ3:} \clearreq{When the \clearvar{absolute\_altitude\_error} becomes either less than 180 ft or greater than 1020 ft, the Altitude Alerting System shall terminate all \clearvar{altitude\_alerts} within 1 second.}

\small
\vspace{-0.15cm}
\begin{multicols}{2}
\begin{minted}[linenos,xleftmargin=.5cm]{scasp.py:SCASPLexer -x}
happens(term_all_alerts, T) :- 
error_outside(180,1020,T1),T1#<T,T #<T1+1.
initiates(term_alerts,dep_alert_on,T).
terminates(term_alerts,dep_alert_off,T).
initiates(term_alerts,cap_alert_on,T).
terminates(term_alerts,cap_alert_off,T).
initiates(term_alerts,alt_alert_on,T).
terminates(term_alerts,alt_alert_off,T).
\end{minted}
\end{multicols}
\vspace{-0.15cm}
\normalsize

\noindent\textbf{REQ4:} \textit{While the \clearvar{altitude\_selection\_knob} has not been \clearstate{adjusting} within the previous 5 seconds, the  Altitude Alerting System shall determine \clearvar{absolute\_al\-titude\_error} as the absolute difference between the \clearvar{selected\_altitude} and the \clearvar{barometric\_altitude}, otherwise  \clearvar{absolute\_altitude\_error} shall be 0.}

\small
\vspace{-0.15cm}
\begin{minted}[linenos,xleftmargin=.5cm]{scasp.py:SCASPLexer -x}
holdsAt(altitude_abs_error(E),T) :-
        holdsAt(barometric_alt(B),T), holdsAt(selected_alt(S),T),
        abs(S,B,E),T1 #= T-5, not happensIn(adjust(Value),T1,T).
holdsAt(altitude_abs_error(0), T) :-
        happens(adjust(Value), T1, T1), T #>= T1, T #=< T1 + 5 
happensIn(Event, T1, T2)          :- happens(Event, T), T1 #=< T, T #=< T2.
\end{minted}
\vspace{-0.15cm}
\normalsize

\noindent\textbf{REQ5:} \textit{Upon initialization, the Altitude Alerting System shall consider the \clearvar{altitude\_selection\_knob} to be \clearstate{adjusting}.}


\small
\vspace{-0.15cm}
\begin{multicols}{2}
\begin{minted}[linenos,xleftmargin=.5cm]{scasp.py:SCASPLexer -x}
initiallyP(altitude_adjusting).
initiates(reset, altitude_adjusting, T).
\end{minted}
\end{multicols}
\vspace{-0.15cm}
\normalsize

\subsection{Simulation of Alerting System Runs from Initial Conditions} 
The coding of the system requirements in event calculus and ASP is a logic program that can be executed on the s(CASP) system. We can run various queries to check if certain properties (e.g., safety properties) that we expect to hold, are entailed by the encoding. We can construct various scenarios (narratives, in event calculus parlance) and check that our requirements specifications are indeed consistent, i.e., when we do simulation runs, we do obtain a model. For example, we assume the following initial state of the system and compute when the alerts will be raised ({\tt initiallyP(F)} means that initially fluent {\tt F} is true and {\tt initiallyN(F)} means initially fluent {\tt F} is false). The fluent \texttt{start} is used to define the rate of change of barometric altitude as an event calculus trajectory\footnote{We have given encoding of trajectory in Basic EC, the encoding of trajectory changes only slightly in Extended EC}.   

 With appropriate narratives\footnote{We excluded
 \mintinline{scasp.py:SCASPLexer -x}{initiallyN(cap_alert_on), initiallyN(dep_alert_on),
 initiallyN(alt_alert_on)}  for lack of space as part of the narrative given} for the altitude alerting system, its end-to-end behavior can be understood. To find out the time at which {\tt departure\_alert\_on} fluent holds, we issue the query Q1: \mintinline{scasp.py:SCASPLexer -x}{?- holdsAt(dep_alert_on, T).}
which yields the answer \texttt{T > 30}. This is because the altitude error becomes greater than 200 ft at \texttt{T = 30}.

\small
\vspace{-0.15cm}
\begin{tabular}{p{6cm}p{6cm}}
\begin{minted}[linenos,xleftmargin=.2cm]{scasp.py:SCASPLexer -x}
initiallyP(cap_alert_off).           
initiallyP(dep_alert_off).         
initiallyP(alt_alert_off).               
initiallyP(altitude_adjusting).           
\end{minted}
&
\begin{minted}[linenos,xleftmargin=.2cm,firstnumber=last]{scasp.py:SCASPLexer -x}
initiallyP(selected_alt(32000)).         
initiallyN(selected_alt(V)) :- V \= 32000.        
trajectory(start,T1,barometric_alt(B),T2) :- 
     B #= 32200 + (T2 - T1) * 10.   
\end{minted}
\end{tabular}
\vspace{-0.15cm}
\normalsize


\subsection{Adequacy of Requirements using Knowledge of Environmental Upsets}
We can also test the behavior of the above scenario augmented with domain knowledge. Our goal is test the adequacy of the requirements using prior knowledge of environmental upsets. Adequacy means that in the presence of known failures, the requirements are still consistent, i.e., a model exists for them.
In this case, the knowledge is simple and relates to Single Event Upsets (or SEUs)~\cite{normand1996single} which can induce random resets of avionic software systems. We can model the destructive power of the SEU using a \textit{reset} event. The \textit{reset} event overrides the constructed internal state of the system, forcing the initialization state to be re-established.
Our goal in introducing this reset is to `move the designer to consider how robust the system initialization logic is to such transient resets.
In ASP, an extraneous event (e.g., reset) is represented as an abducible (a reset may or may not happen). We would want to know if our system will still behave correctly in presence of this extraneous event. The knowledge assumes that the reset event can only override the cyber system (software state) and does not affect the continuous state of the physical world. In case of the altitude alerting system, the reset event can only affect the alert status and the altitude adjusting status, but cannot affect the barometric altitude of the aircraft or the selected altitude value. This in turn maps to the reset event terminating only the fluents that describe the internal software state of the system. The reset will not terminate fluents that describe the state evolution of continuous quantities involving the real world. In general, fluents that are outside the scope software system boundary are not affected by reset. 
The following generic rule specifies that the reset event terminates every fluent associated with the cyber system and initiates it to an initial value.

\vspace{-0.3cm}
\small
\begin{minted}[linenos,xleftmargin=.5cm]{scasp.py:SCASPLexer -x}
terminates(reset, Fluent, T)  :- internal_state(Fluent).
initiates(reset, Default, T)  :- default(Fluent, Default), internal_state(Fluent).
\end{minted}
\normalsize
\vspace{-0.3cm}

\textbf{Illustrating a Failing Run:} 
Let the predicate $in\_alerting\_zone(T)$ denote that the altitude error conditions for altitude alert ($REQ2$) are satisfied. We show that the system fails to issue critical alerts in the presence of an SEU-induced reset.  
We can check whether the altitude alerts are missed while in an alerting zone using the query \mintinline{scasp.py:SCASPLexer -x}{?- in_alerting_zone(T), holdsAt(altitude_alert_off, T)} . With correct assumptions, the query would return `no models' in s(CASP).  
By assuming an incorrect requirement in place of $REQ5$, the above query returns a binding in s(CASP), thereby showing a missed alert.   The incorrect requirement, $REQ5'$, is stated as: \textit{Upon initialization, the Altitude Alerting System shall consider the \clearvar{altitude\_selection\_knob} to be \textbf{not} \clearstate{Adjusting}}. More than likely, $REQ5$ may not even be stated by a system designer, and if stated, it is likely that it will be in the form of $REQ5'$, as one would \textit{assume} that the pilot would not worry about adjusting the altitude knob during plane's takeoff. 

Assume that the plane is cruising at 32,200 feet at time $t$ and selected altitude value is at 32300 ft. The plane's barometric altitude increases at the rate of 10 ft per second. At $t+30$, the altitude error becomes greater than 200 ft. Normally, the system should issue altitude alert at $t+30$.  However, a reset occurs at $t+30$. The reset sets the altitude to not adjusting as per $REQ5'$ and also turns the altitude alert off. Due to $REQ5'$, the altitude is treated as not adjusting for any time $T > t+30$. Thus, although   $in\_alerting\_zone(T)$ is true for $T > t+30$, the departure alert is never turned back on. Therefore, an alert that should be issued to the pilot is missed by the system. 
On the other hand, when $REQ5$ is used, the altitude error is forced to be zero upon reset. This ensures that at $T = t+35$, (5 seconds from $t+30$) the error becomes 200 ft, thereby making $in\_alerting\_zone(t+35)$ true. Again, per $REQ2$, the altitude alert would be issued and no alert is missed. 
Adequacy of requirements with reset is checked easily using query (Q2):
\mintinline{scasp.py:SCASPLexer -x}{?-happens(reset,T),T1#>=T+5,in_alerting_zone(T1),holdsAt(alt_alert_off,T2),T2#>T1.
}

\mycomment{
\small
\begin{minted}{scasp.py:SCASPLexer -x}
?- happens(reset,T), T1 #>= T+5, in_alerting_zone(T1), holdsAt(alt_alert_off,T2), T2 #> T1.
\end{minted}
\normalsize
}

The above query (Q2) asks s(CASP), that if in the event of a reset, is the altitude alert turned on after 5 seconds? 
If Q2 succeeds, then the requirements are adequate. If Q2 fails (say when we use $REQ5$ instead of $REQ5'$) then the requirements are not adequate. This example shows the ease of validating requirements using EC and s(CASP) while using knowledge of SEUs.
\mycomment{Another important point to note is despite a reset event overriding the internal state of the cyber system, the real world fluents such as barometric altitude continue unabated. This is directly modelled through EC's \textit{inertia} axioms.} In summary, our ASP-coded EC model of the requirements are used to check for consistency and that they satisfy certain properties much in the spirit of model checking. \textit{This ASP and EC model can also be elegantly used for checking adequacy of requirements against known failure patterns. The failure patterns can be induced into the model through the use of abduction directly supported in s(CASP)}. This is possible due to near-zero ``semantic gap" between the requirements and their mapping to ASP and EC. Checking adequacy is not direct in other approaches based on automata and Kripke structures. 

\subsection{Performance}

The performance of s(CASP) for above queries is given in Table 1. The queries were run on a quad code Intel i7-10510U processor at 1.8Ghz and 8GB RAM. For comparison, we also attempted to run the queries on CLINGO \cite{clingo} that is based on grounding and SAT-solving. Due to large size of grounded program produced from discretization of time and altitude values, the queries generally out of memory. Note that Q3 is Q2 but rerun with REQ5 replaced by REQ5'. 

\begin{table}
\label{tab:performance}
\small
\centering
\begin{tabular}{|c|c|c|}
\hline
\textbf{Query} & \textbf{s(CASP)} & \textbf{CLINGO} \\
\hline
Query Q1, Normal Run & 0.930s & > 40 min \\
\hline
Query Q2, With Reset & 2.077s & > 40 min \\
\hline
Query Q3 With Reset & 0.081s & > 40 min \\
\hline
\end{tabular}
 \caption{Query Execution Times}
\end{table}
\normalsize
\par\noindent

\section{Conclusions and Future Work}
\label{concl}

This paper presents a novel, systematic approach to formalizing requirement modeling, which enables the integration and leveraging of domain knowledge toward a computer-aided system validation, verification and assurance. We see this work as a first step towards a new generation of  Knowledge-Assisted System Engineering (KASE) tooling. Our goal with this class of tooling is to significantly reduce system development costs by systematically addressing requirements specification defects much earlier in the development process.  Our approach is unique. It enables a formal examination of preliminary conceptual models of intended behaviours, to assess adequacy and consistency with respect to world assumptions, anticipated failure modes, and environmental conditions. Our approach further enables the application of domain knowledge and prior lessons learned  (known deficiency areas) to be fused with emerging specifications to more rapidly mature design robustness.

The primary catalyst for our success is not only the simplicity and elegance of the EC, ASP, and the s(CASP) system,  which allows us to approach human levels of reasoning, it is also the intuitive formalism of EC being hidden behind the CLEAR natural language style of specification, that enables the power of formal methods to be introduced with minimal training and required specialist knowledge. We hope this will remove some of the barriers to broader spread of formal method adoption \cite{davis2013study}.

To the best of our knowledge, the work reported here is novel, as the event calculus has not been used outside of AI. Its application to system assurance here is novel. Even within AI, EC has not been widely adopted because of the lack of sound, query-driven implementations of negation-as-failure that can handle time faithfully as a continuous quantity. Our s(CASP) implementation of ASP solves both these problems. The work reported here also has practical value, as its results have benefited engineering teams at a major aerospace company.
Compared to model checking techniques, our models are closer to the requirement specification, as states do not have to be discerned in advance. Moreover, reasoning over continuous time in EC is realized directly in the s(CASP) system. 


There are many potential avenues for follow-on areas of research.  First, we have already internally established the feasibility to automatically generate the ASP formal model from the MIDAS model and CLEAR requirements, using various methods, including simple definite clause grammars to more state-of-the-art symbolic natural language processing approaches~\cite{basu2021aaai}. We are also exploring formal architectural reasoning using models of defined functional intent and architectural intent with the abductive inference of component failures, as well as modeling more complex cyber physical systems using the method reported in this paper. Finally, we plan to apply our techniques to requirements specification analysis of more systems, in a manner similar to UPPAAL (\url{https://uppaal.org/casestudies/}).

\mycomment{
\medskip\noindent{\bf Acknowledgement:}
UT Dallas authors are supported by NSF (grants IIS 1718945, IIS 1910131, IIP 1916206), Amazon and DoD. Support for this project also comes from the FIT BUT internal project FIT-S-20-6427 and by the H2020 ECSEL project Arrowhead Tools.
}

\bibliographystyle{eptcs}
\bibliography{references}

\begin{thebibliography}{10}
\providecommand{\bibitemdeclare}[2]{}
\providecommand{\surnamestart}{}
\providecommand{\surnameend}{}
\providecommand{\urlprefix}{Available at }
\providecommand{\url}[1]{\texttt{#1}}
\providecommand{\href}[2]{\texttt{#2}}
\providecommand{\urlalt}[2]{\href{#1}{#2}}
\providecommand{\doi}[1]{doi:\urlalt{http://dx.doi.org/#1}{#1}}
\providecommand{\bibinfo}[2]{#2}

\bibitemdeclare{inproceedings}{basu2021aaai}
\bibitem{basu2021aaai}
\bibinfo{author}{Kinjal~Basu \surnamestart et. al\surnameend}
  (\bibinfo{year}{2021}): \emph{\bibinfo{title}{Knowledge-driven Natural
  Language Understanding of English Text and its Applications}}.
\newblock In: {\sl \bibinfo{booktitle}{Thirty-Fifth {AAAI} Conference on
  Artificial Intelligence, {AAAI} 2021, 2021}}, \bibinfo{publisher}{{AAAI}
  Press}, pp. \bibinfo{pages}{12554--12563}.

\bibitemdeclare{article}{arias2018constraint}
\bibitem{arias2018constraint}
\bibinfo{author}{Joaqu{\'i}n \surnamestart Arias\surnameend},
  \bibinfo{author}{Manuel \surnamestart Carro\surnameend},
  \bibinfo{author}{Elmer \surnamestart Salazar\surnameend},
  \bibinfo{author}{Kyle \surnamestart Marple\surnameend} \&
  \bibinfo{author}{Gopal \surnamestart Gupta\surnameend}
  (\bibinfo{year}{2018}): \emph{\bibinfo{title}{Constraint answer set
  programming without grounding}}.
\newblock {\sl \bibinfo{journal}{TPLP}} \bibinfo{volume}{18(3-4):337-354},
  \doi{10.1017/S1471068418000285}.

\bibitemdeclare{inproceedings}{arias2019modeling-short}
\bibitem{arias2019modeling-short}
\bibinfo{author}{Joaqu{\'i}n et~al. \surnamestart Arias\surnameend}
  (\bibinfo{year}{2019}): \emph{\bibinfo{title}{Modeling and Reasoning in Event
  Calculus Using Goal-Directed Constraint Answer Set Programming}}.
\newblock In: {\sl \bibinfo{booktitle}{LOPSTR'19}},
  \bibinfo{organization}{Springer}, pp. \bibinfo{pages}{139--155},
  \doi{10.1007/978-3-030-45260-5\_9}.

\bibitemdeclare{book}{baral}
\bibitem{baral}
\bibinfo{author}{C.~\surnamestart Baral\surnameend} (\bibinfo{year}{2003}):
  \emph{\bibinfo{title}{Knowledge representation, reasoning and declarative
  problem solving}}.
\newblock \bibinfo{publisher}{Cambridge University Press},
  \doi{10.1017/CBO9780511543357}.

\bibitemdeclare{inproceedings}{bhatt2018clearfide}
\bibitem{bhatt2018clearfide}
\bibinfo{author}{Devesh \surnamestart Bhatt\surnameend},
  \bibinfo{author}{Brendan \surnamestart Hall\surnameend} et~al.
  (\bibinfo{year}{2018}): \emph{\bibinfo{title}{The CLEAR Way To Transparent
  Formal Methods}}.
\newblock In: {\sl \bibinfo{booktitle}{4th Workshop on Formal Integrated
  Development Environment (F-IDE), FLoC 2018}}.

\bibitemdeclare{inproceedings}{davis2013study}
\bibitem{davis2013study}
\bibinfo{author}{Jennifer \surnamestart Davis\surnameend} et~al.
  (\bibinfo{year}{2013}): \emph{\bibinfo{title}{Study on the barriers to the
  industrial adoption of formal methods}}.
\newblock In: {\sl \bibinfo{booktitle}{International Workshop on Formal Methods
  for Industrial Critical Systems}}, \bibinfo{organization}{Springer}, pp.
  \bibinfo{pages}{63--77}, \doi{10.1007/978-3-642-41010-9\_5}.

\bibitemdeclare{incollection}{dori2011object}
\bibitem{dori2011object}
\bibinfo{author}{Dov \surnamestart Dori\surnameend} (\bibinfo{year}{2011}):
  \emph{\bibinfo{title}{Object-process methodology}}.
\newblock In: {\sl \bibinfo{booktitle}{Encyclopedia of Knowledge Management,
  Second Edition}}, \bibinfo{publisher}{IGI Global}, pp.
  \bibinfo{pages}{1208--1220}, \doi{10.4018/978-1-59904-931-1.ch116}.

\bibitemdeclare{article}{clingo}
\bibitem{clingo}
\bibinfo{author}{Martin \surnamestart Gebser\surnameend} et~al.
  (\bibinfo{year}{2011}): \emph{\bibinfo{title}{Potassco: The Potsdam answer
  set solving collection}}.
\newblock {\sl \bibinfo{journal}{Ai Communications}}
  \bibinfo{volume}{24}(\bibinfo{number}{2}), pp. \bibinfo{pages}{107--124},
  \doi{10.3233/AIC-2011-0491}.

\bibitemdeclare{book}{gelfond2014knowledge}
\bibitem{gelfond2014knowledge}
\bibinfo{author}{M.~\surnamestart Gelfond\surnameend} \&
  \bibinfo{author}{Y.~\surnamestart Kahl\surnameend} (\bibinfo{year}{2014}):
  \emph{\bibinfo{title}{Knowledge representation, reasoning, \& design of
  intelligent agents: The answer-set programming approach}}.
\newblock \bibinfo{publisher}{Cambridge Univ. Press},
  \doi{10.1017/CBO9781139342124}.

\bibitemdeclare{inproceedings}{hall2018clear}
\bibitem{hall2018clear}
\bibinfo{author}{B.~\surnamestart Hall\surnameend},
  \bibinfo{author}{D.~\surnamestart Bhatt\surnameend} et~al.
  (\bibinfo{year}{2018}): \emph{\bibinfo{title}{A CLEAR Adoption of EARS}}.
\newblock In: {\sl \bibinfo{booktitle}{IEEE EARS Workshop}}, pp.
  \bibinfo{pages}{14--15}, \doi{10.1109/EARS.2018.00010}.

\bibitemdeclare{inproceedings}{hall2020model}
\bibitem{hall2020model}
\bibinfo{author}{B.~\surnamestart Hall\surnameend},
  \bibinfo{author}{J.~\surnamestart Fiedor\surnameend} \&
  \bibinfo{author}{Y~\surnamestart Jeppu\surnameend} (\bibinfo{year}{2020}):
  \emph{\bibinfo{title}{Model Integrated Decomposition and Assisted
  Specification (MIDAS)}}.
\newblock In: {\sl \bibinfo{booktitle}{INCOSE Int'l Symp.}},
  \bibinfo{volume}{30(1)}, \bibinfo{organization}{Wiley}, pp.
  \bibinfo{pages}{821--841}.

\bibitemdeclare{article}{harman65}
\bibitem{harman65}
\bibinfo{author}{G.~H. \surnamestart Harman\surnameend} (\bibinfo{year}{1965}):
  \emph{\bibinfo{title}{The Inference to the Best Explanation}}.
\newblock {\sl \bibinfo{journal}{The Philosophical Review}}
  \bibinfo{volume}{74}(\bibinfo{number}{1}), pp. \bibinfo{pages}{88--95},
  \doi{10.2307/2183532}.

\bibitemdeclare{inproceedings}{mavin2010big}
\bibitem{mavin2010big}
\bibinfo{author}{Alistair \surnamestart Mavin\surnameend} \&
  \bibinfo{author}{Philip \surnamestart Wilkinson\surnameend}
  (\bibinfo{year}{2010}): \emph{\bibinfo{title}{Big ears (the return of" easy
  approach to requirements engineering")}}.
\newblock In: {\sl \bibinfo{booktitle}{2010 18th IEEE International
  Requirements Engineering Conference}}, \bibinfo{organization}{IEEE}, pp.
  \bibinfo{pages}{277--282}, \doi{10.1109/RE.2010.39}.

\bibitemdeclare{inproceedings}{mavin2009easy}
\bibitem{mavin2009easy}
\bibinfo{author}{Alistair \surnamestart Mavin\surnameend} et~al.
  (\bibinfo{year}{2009}): \emph{\bibinfo{title}{Easy approach to requirements
  syntax (EARS)}}.
\newblock In: {\sl \bibinfo{booktitle}{2009 17th IEEE International
  Requirements Engineering Conference}}, \bibinfo{organization}{IEEE}, pp.
  \bibinfo{pages}{317--322}, \doi{10.1109/RE.2009.9}.

\bibitemdeclare{inproceedings}{mavin2016listens}
\bibitem{mavin2016listens}
\bibinfo{author}{Alistair \surnamestart Mavin\surnameend} et~al.
  (\bibinfo{year}{2016}): \emph{\bibinfo{title}{Listens learned (8 lessons
  learned applying EARS)}}.
\newblock In: {\sl \bibinfo{booktitle}{2016 IEEE 24th International
  Requirements Engineering Conference (RE)}}, \bibinfo{organization}{IEEE}, pp.
  \bibinfo{pages}{276--282}, \doi{10.1109/RE.2016.38}.

\bibitemdeclare{book}{micouin2014model}
\bibitem{micouin2014model}
\bibinfo{author}{Patrice \surnamestart Micouin\surnameend}
  (\bibinfo{year}{2014}): \emph{\bibinfo{title}{Model Based Systems
  Engineering: Fundamentals and Methods}}.
\newblock \bibinfo{publisher}{John Wiley \& Sons}, \doi{10.1002/9781118579435}.

\bibitemdeclare{book}{mueller_book}
\bibitem{mueller_book}
\bibinfo{author}{Erik~T. \surnamestart Mueller\surnameend}
  (\bibinfo{year}{2014}): \emph{\bibinfo{title}{Common Sense Reasoning: An
  Event Calculus Based Approach (2nd Edition)}}.
\newblock \bibinfo{publisher}{Morgan Kaufmann}.

\bibitemdeclare{article}{normand1996single}
\bibitem{normand1996single}
\bibinfo{author}{Eugene \surnamestart Normand\surnameend}
  (\bibinfo{year}{1996}): \emph{\bibinfo{title}{Single-event effects in
  avionics}}.
\newblock {\sl \bibinfo{journal}{IEEE Transactions on nuclear science}}
  \bibinfo{volume}{43}(\bibinfo{number}{2}), pp. \bibinfo{pages}{461--474},
  \doi{10.1109/23.490893}.

\bibitemdeclare{article}{sergot1986logic}
\bibitem{sergot1986logic}
\bibinfo{author}{M~\surnamestart Sergot\surnameend} \&
  \bibinfo{author}{R~\surnamestart Kowalski\surnameend} (\bibinfo{year}{1986}):
  \emph{\bibinfo{title}{A logic-based calculus of events}}.
\newblock {\sl \bibinfo{journal}{New Generation Computing}}
  \bibinfo{volume}{4}(\bibinfo{number}{1}), pp. \bibinfo{pages}{67--95},
  \doi{10.1007/BF03037383}.

\bibitemdeclare{incollection}{shanahan1999event}
\bibitem{shanahan1999event}
\bibinfo{author}{Murray \surnamestart Shanahan\surnameend}
  (\bibinfo{year}{1999}): \emph{\bibinfo{title}{The event calculus explained}}.
\newblock In: {\sl \bibinfo{booktitle}{Artificial intelligence today}},
  \bibinfo{publisher}{Springer}, pp. \bibinfo{pages}{409--430},
  \doi{10.1007/3-540-48317-9\_17}.

\bibitemdeclare{misc}{a400mseville}
\bibitem{a400mseville}
\bibinfo{author}{\surnamestart WikePedia\surnameend}:
  \emph{\bibinfo{title}{2015 Seville {A}irbus {A400M} crash}}.
\newblock
  \bibinfo{howpublished}{\url{https://en.wikipedia.org/wiki/2015_Seville_Airbus_A400M_crash}}.

\end{thebibliography}

\end{document}